\documentclass[%
 reprint,
 amsmath,amssymb,
 longbibliography,
 aps,
prb,
]{revtex4-2}

\usepackage{graphicx}%
\usepackage{dcolumn}%
\usepackage{bm}%
\usepackage{hyperref}%

\usepackage{etoolbox}
\makeatletter
\patchcmd\linenumberpar{\@LN@parpgbrk}{\penalty\@LN@parpgpen\relax}{}{}
\makeatother

\usepackage{xcolor}
\usepackage{braket}
\usepackage[caption=false]{subfig}

\newcommand{\COMMENTED}[1]{}
\newcommand*{\colorboxed}{}
\def\colorboxed#1#{%
  \colorboxedAux{#1}%
}
\newcommand*{\colorboxedAux}[3]{%
  \begingroup
    \colorlet{cb@saved}{.}%
    \color#1{#2}%
    \boxed{%
      \color{cb@saved}%
      #3%
    }%
  \endgroup
}

\begin{document}

\title{Automatic Order Detection and Restoration Through Systematically Improvable Variational Wave Functions}

\author{Ryan Levy}
 \email{rlevy@flatironinstitute.org}
\author{Miguel A. Morales}%
\author{Shiwei Zhang}
 
\affiliation{%
Center for Computational Quantum Physics, Flatiron Institute, New York, NY, 10010, USA
}%

\date{\today}%

\begin{abstract}
Variational wave function ansatze are an invaluable tool to study the
properties of strongly correlated systems. We propose such a wave function, 
based on the theory of auxiliary fields and combining aspects
of auxiliary-field quantum Monte Carlo and 
modern variational optimization techniques including automatic differentiation.
The resulting ansatz,   
consisting of several slices of optimized projectors,
is highly expressive and systematically improvable. 
We benchmark this form on 
the two-dimensional  Hubbard model, using both cylindrical and large, fully periodic supercells. 
The computed ground-state energies are competitive with the best variational results. Moreover, the optimized wave functions predict the correct 
ground-state order with near full symmetry  
restoration (i.e. translation invariance) despite initial states with incorrect orders.  
The ansatz can become a tool for
local order prediction, leading to a new paradigm for variational studies of bulk systems.
It can also be viewed as an approach to produce accurate and systematically improvable wave functions in a convenient form of  non-orthogonal Slater determinants 
(e.g., for quantum chemistry)
at polynomial computational cost. 
\end{abstract}

\maketitle

\section{Introduction}
Understanding the ground state properties of strongly correlated systems has been an important goal in the study of electronic systems. There has been much success in utilizing variational methods, such as  
variational Monte Carlo (VMC) \cite{PhysRev.138.A442,Ceperley1977} 
and density matrix renormalization group (DMRG) \cite{White1992,White1993,Schollwck2011}, however each has
drawbacks. Tensor network methods, while having incredible variational power, often have challenges of converging large width cylinders or periodic boundary conditions with finite resources \cite{Stoudenmire2012}. In order to understand these large bulk systems, quantum Monte Carlo (QMC) methods \cite{Ceperley1986,becca2017quantum} are invaluable, however interesting regimes can be plagued with a sign problem, which renders many variational calculations exponentially difficult. Mitigation of this problem, for example using constrained path auxiliary field QMC (CP-AFQMC) \cite{Zhang1997}, can provide accurate local energies but require more sophisticated methods, such as back-propagation, to measure observables that do not commute with the Hamiltonian. 
 
In VMC, a parameterized wave function ansatz $\ket{\psi}$ is optimized against the variational energy expectation $\langle E\rangle = \braket{\psi|\hat{H}|\psi}/\braket{\psi|\psi}$. For many common ansatze, this evaluation takes place by stochastic sampling via Monte Carlo in an occupation basis.  
The key to this approach is the interplay between expressivity of the ansatz, and the ability to optimize it to a global minimum. 
Often the limiting factor has been that the
commonly used forms for fermionic systems lack sufficient expressivity
to capture different quantum phases under the same wave function ansatz,
thereby making it difficult to detect orders in an unbiased manner. 

Recently there has been a number of fermionic wave functions parameterizations introduced that are inspired from machine learning algorithms \cite{Nomura2017,Luo2019,RobledoMoreno2022,Autoregressive}. These are generally systematically improvable and 
have been shown to obtain very close agreement with the true ground state in small 
to medium-sized systems. 
However, 
in 
order to optimize vast numbers of parameters, the state may need to be constrained to certain physics, e.g. superconductivity or the wavelength of a charge density wave \cite{RobledoMoreno2022,varbench}. 
Furthermore, a wave function ansatz with 
many parameters can make it more challenging to connect with the underlying physics. In strongly correlated systems with competing orders 
being separated by small energy differences, the variational energy 
can cease being an effective signal for detection of different phases.
An  ansatz with a smaller number of parameters and a simpler, more
physically-inspired construction can be advantageous. 

\begin{figure*}[t]
    \centering
    \includegraphics[width=2\columnwidth]{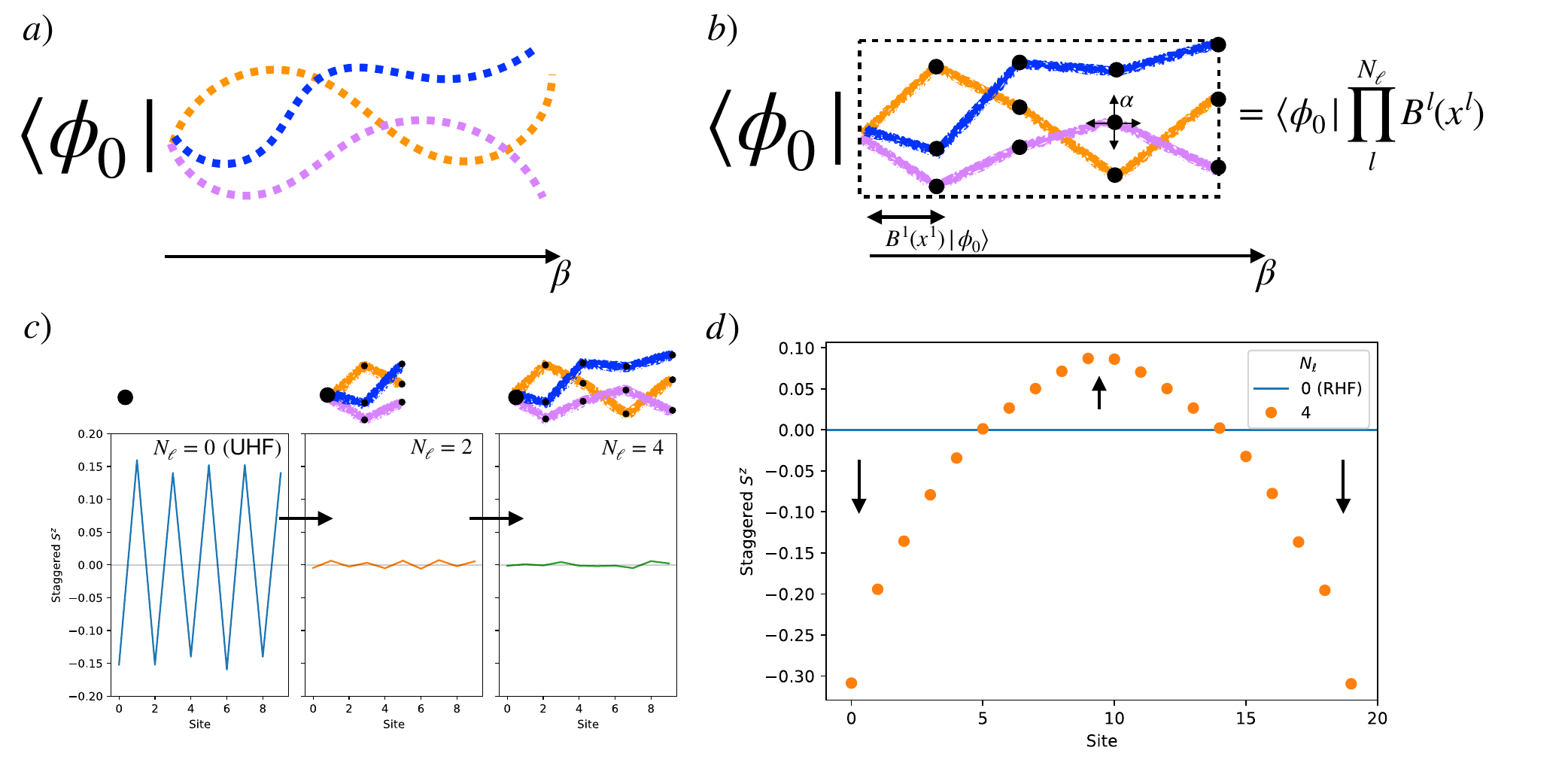}
    \caption{a) Cartoon of 
    standard 
    AFQMC where random walks are taken with many
    small steps of size $\tau$ , 
    following a pre-defined projector, %
    to reach an imaginary time of $\beta$. 
    b) cartoon of VAFQMC random walk, where a  number of potentially large steps are taken,
    following an {\it effective\/} projector that is variationally optimized,
    to reach the equivalent imaginary time of $\beta$. Each dot represents a projector slice and is a function of optimizable parameters $\alpha$. 
    c) Staggered magnetization measurement for the 3rd column of a $10\times 10$ $U=4$ Hubbard model with $n=0.8$ density. The first pane shows the initial unrestricted Hartree-Fock (UHF) solution, and subsequent panes show VAFQMC  slowly projecting out the incorrect initial order to create a spatially uniform ground state. 
    d) Staggered magnetization for a $20 
    \times 4$ Hubbard model at $U=4$ and $n=0.9$ density,
    with magnetic pinning fields applied at the edges. Starting from a 
    qualitatively incorrect 
    RHF initial state and optimizing the projection can uncover the correct ground state local order. }
    \label{fig:cartoon}
\end{figure*}
We present a variational wave functions ansatz based on the theory of auxiliary fields,
which aims to have ${\mathcal O}(N^2)$ or less 
optimizable variational parameters and yields a simple form of 
the wave function that is straightforward to systematically improve. 
The ansatz builds on a form 
originally introduced in Ref.~\cite{sorellaVAFQMC} by 
Sorella, 
as a 
modified version of auxiliary field quantum Monte Carlo (AFQMC), and given the name variational AFQMC, or VAFQMC. 
We introduce a further generalized 
variational form, which defines a
framework with which the expressivity can be easily expanded.
An extension was recently presented to treat molecules in Ref.~\cite{chenHAFQMC},
with a different form than what we consider. 
We benchmark our variational wave function in the Hubbard model against existing VMC results, as well as DMRG and two 
projector QMC methods: AFQMC \cite{Zhang1997,Shi2013} and fixed-node diffusion Monte Carlo \cite{LeBlanc2015,varbench}. 

The rest of the paper is organized as follows. 
In section~\ref{sec:methods}, we introduce our variational formulation and describe the implementation details. Then in section~\ref{sec:results} we benchmark against state of the art VMC, projector QMC, and tensor network methods, showing competitive variational energies and predictive power of local observables. Finally in section~\ref{sec:conclusion} we 
conclude and discuss the 
outlook 
of VAFQMC in conjunction with other methods.

\section{Methods}\label{sec:methods}

We consider a variational form modeled after the projector method \cite{Trivedi1990} such that the ground state $\ket{\psi_0}$ of a Hamiltonian $H$ can be obtained via
\begin{equation}
    \ket{\psi_0} = \lim_{N\to\infty}(e^{-\tau H})^N \ket{\psi_T},
\end{equation}
for some trial state $\ket{\psi_T}$ which has non-zero overlap with the ground state. Traditionally, as in AFQMC, a stochastic random walk is formulated to project the trial wave function into the ground state via a Monte Carlo routine. 
A cartoon representation of this process is depicted in Fig.~\ref{fig:cartoon}a. Instead, by fixing the number of projection steps and optimizing the projectors as a wave function ansatz, 
we can produce an integral representation of the variational state shown in Fig.~\ref{fig:cartoon}b.

Our formalism will apply to all Hamiltonians containing general one-body and two-body terms. For concreteness we will use the Hubbard model \cite{Hubbard1963} in this work to both describe our approach and perform 
benchmark tests. The Hubbard model is given by
\begin{equation}
    \hat{H} = \hat{T} + \hat{V} = -t\sum_{\langle ij\rangle\sigma } c^\dagger_{i\sigma}c_{j\sigma} + h.c. + U \sum_i n_{\uparrow i}n_{\downarrow i},
\end{equation}
where $\langle ij\rangle$ denotes nearest-neighbor hopping,
$U/t$ is the strength of the on-site interaction. 
We restrict the model to 2D such that $i=(i_x,i_y)$,
and consider lattices of size $N_x\times N_y$. Our focus is on 
describing ground-state order and approaching the 2D limit. 
For this we study both cylindrical geometry, with 
open boundary condition along $x$ and periodic boundary condition (PBC) along $y$, and square lattices with fully PBCs along both direction. 
The cylindrical cells allow us to perform systematic benchmarks on 
the description of spin and charge density waves, the primary order in the
ground state of this model. The fully periodic supercells allow us to test 
the quality of the wave function --- including symmetry restoration --- and scaling as we approach the 2D bulk limit. 

To apply the projection operator $\exp(-\tau \hat{H})$ to the trial wave function, in AFQMC the propogator is split up into a kinetic and potential part $\exp(-\tau \hat{H)}\approx \exp(-\tau \hat{T})\exp(-\tau \hat{V})$. 
This step incurs an error of ${\mathcal O}(\tau)$ and must be corrected, but this is not of concern here due to the variational nature of our wave function as discussed below. 
The kinetic term can be applied to the trial wave function as a transformation of the Slater determinant, via Thouless theorem \cite{Thouless1960}. For the potential term, a 
 Hubbard-Stratonovich (HS) decomposition can be applied \cite{stratonovich1957method,Hirsch1983}, for example the discrete version 
\begin{equation}
    e^{-\tau U n_{\uparrow i}n_{\downarrow i}} \propto \sum_{x_i = \pm 1} \exp(\lambda_i x_i (n_{\uparrow i}-n_{\downarrow i})),
\end{equation}
where $\cosh \lambda_i = \exp(U\tau)$. 

The form of our variational wave function can be thought of as a sum over auxiliary fields: 
\begin{equation}
    |\psi\rangle = \sum_{x^l} \prod_l^{N_\ell}  e^{\sum_i \lambda_i x_i^l (n_{\uparrow i}-n_{\downarrow i})} e^{-\sum_\sigma \tau^l_\sigma T^l_\sigma }\ket{\phi_0}, %
\label{eq:projector}
\end{equation}
where $T^l_\sigma = t^{l\sigma}_{ij} c^\dagger_{\sigma i}c_{\sigma j}$ and constant terms are ignored.

This can be recast into an {\it effective Hamiltonian\/} form as
\begin{equation}
    \ket{\psi} = \prod_l^{N_\ell} e^{-\tau_l \hat{\mathcal{H}}_l}\ket{\phi_0},
    \label{eq:effectiveH}
\end{equation}
where $\hat{\mathcal{H}}_l$ is determined by optimizable parameters %
$T^l_\sigma$, $\tau_l$, 
as well as the form of the HS,
all of which can be different from that of the actual Hamiltonian $H$. 
In other words, we consider variational wave functions of the form
\begin{equation}
    |\psi\rangle = \int \prod_l^{N_\ell} B^l(x^l)\,dx^l\, \ket{\phi_0} 
    =\sum_{\{x\}} \ket{x} 
\label{eq:wf-form}
\end{equation}
where $B^l(x^l)$ is a one-body propagator of the form $B^l(x^l;\alpha)=\exp[\sum_{ij} \sum_{\sigma\sigma'} h^l_{i\sigma;j\sigma'}(x^l_{ij};\{\alpha\}) c_{i,\sigma}^\dagger c_{j,\sigma'}]$, with $i$ denoting lattice sites (or basis functions), $\ket{x}$ is the resulting application of $\prod B^l(x)\ket{\phi_0}$, and $\sigma$ 
($\sigma'$) denoting spins. The {\it effective\/} one-body ``hamiltonian" in the 
exponent in $B^l$ contains optimization parameters, denoted by $\alpha$. 
For the Hubbard model example we consider here, 
$B^l(x^l)$ is of the simpler, product form in Eq.~(\ref{eq:projector}), with
$\alpha$ 
consisting of $\alpha=\{\lambda_i, \tau^l_\sigma, t_{ij}^{l\sigma}\}$. 
Here we have separated 
$h^l$ in the one-body auxiliary-field-independent term into an overall scaling factor $\tau^l$ and the hopping terms $T^l_{i\sigma;j\sigma}$, so that 
no overall scaling factor
is involved in the optimization of the latter. 
These parameters are unconstrained: $\lambda_i$ is repeated for each projector slice $l$ but can be non-uniform over sites $i$; $\tau^l_\sigma$ is now distinct from $\lambda_i$ and is allowed to be positive or negative; and $T^l$ is no longer constrained to be Hermitian. In practice the values of these parameters are optimized directly. 
Most of these parameters are in $T^l_\sigma$, with $\mathcal{O}(N_\ell\cdot N^2)$ parameters. This general variational ansatz, which is restricted to one-body terms, can also be related to shadow wave functions where a kernel of both physical and auxiliary degrees of freedom is parameterized and auxiliary variables are integrated out \cite{Vitiello1988,Reatto1988}. Shadow wave functions however, have historically been used in a particle configuration basis unlike Eq.~(\ref{eq:wf-form}). 

Unlike Ref.~\cite{sorellaVAFQMC}, our variational form has no fixed projection time $\beta^{N_\ell} = \sum_l^{N_\ell} (\tau^l_\uparrow + \tau^l_\downarrow)/2 $. This allows the ansatz to `grow' as needed in projection time, to potentially reach the ground state with few effective projector steps $N_\ell$. In fact, completely decoupling $\tau^l$ from $\lambda_i$ allows for not only cooling, i.e. $\beta^{l+1} =  \beta^l + \tau$, but \textit{intermediate heating} i.e. $\beta^{l+1} = \beta^l - \tau$. We show below that the optimized results indeed takes advantage of this structure.

Given a variational form of Eq.~(\ref{eq:projector}) or generally Eq.~(\ref{eq:wf-form}), we can evaluate the energy via a double integral
\begin{align}
    \langle E \rangle &= \frac{\braket{\psi|\hat{H}|\psi}}{\braket{\psi|\psi}}\\
    &= \frac{\sum_{x x^\prime} \rho(x,x^\prime) E_L(x,x^\prime) S(x,x^\prime)}{\sum_{x x^\prime} \rho(x,x^\prime) S(x,x^\prime) }\\
    &= \frac{\langle E_L S \rangle_\rho}{\langle S \rangle_\rho},\label{eq:energy}
\end{align}
where each $x$ (or $x'$) is a field configuration that produces a Slater determinant $\ket{x}$ ($\ket{x^\prime}$),
$E_L(x,x^\prime) = \braket{x^\prime|\hat{H}|x}/\braket{x^\prime|x}$ which is similar to a local-energy, $S(x,x^\prime) =\braket{x^\prime|x}/|\braket{x^\prime|x}| $ is the phase or sign of the configuration, and $\rho(x,x^\prime) = |\braket{x^\prime|x}|$. The probability distribution $|\braket{x^\prime|x}|$ can be sampled using Markov chain Monte Carlo (MCMC) through conventional Metropolis algorithm methods \cite{becca2017quantum} where each state is a configuration of fields $(x,x^\prime)$. This procedure is distinctly different from VMC however, and is instead closer to the evaluation of a $N_\ell$-slice path integral in imaginary time. Note that because $\braket{x^\prime|x}$ is not necessarily positive semi-definite a sign problem can occur. However, for the problems we consider, 
the number of slices, $N_l$, is modest and the average $\langle S\rangle$ remains positive as further discussed below. 

Energy optimization is carried out through gradient descent. The derivation of the derivative is provided in Ref.~\cite{sorellaVAFQMC} which we repeat here for convenience. The derivative of of parameter $\alpha$ is given by 
\begin{equation}
    \frac{\partial \braket{\hat{H}}_\rho}{\partial \alpha} = \frac{\braket{S \left( \frac{\partial E_L}{\partial \alpha}+(E_L-\braket{\hat{H}})O\right)}_\rho}{\braket{S}_\rho},
    \label{eq:deriv}
\end{equation}
where $O=\frac{\partial \ln (S\rho)}{\partial \alpha}$. Both $\frac{\partial E_L}{\partial \alpha}$ and $O$ can be easily evaluated using automatic differentiation. By sampling $\rho$ and accumulating the values in eq.(~\ref{eq:deriv}), optimization of the wave function can be performed. This optimization procedure of VAFQMC is therefore dominated by both matrix multiplication (the projection of $\ket{\phi_0}$) and determinant calculations for the local energy which implies VAFQMC scales as $\approx O(N_\ell \cdot N^3)$.
We note that standard practices  in AFQMC, for example, compact decomposition of long-range 
interaction (or Jastrow),
the use of 
force bias \cite{Shi2015} or other update schemes, stabilization, low-rank decomposition, fast computations of local energy etc \cite{Motta-WIREs-2018}, can all be applied to this method.

\subsection{Implementation Details}

We consider the standard Hubbard model with only nearest neighbor hopping.
Two types of calculations are considered, using 
two different boundary conditions: fully periodic (PBC) and cylindrical boundary conditions (CBC). When using cylindrical boundaries, we apply a weak edge pinning field  $H_{pin} =  
\sum_{i\in \text{edge}} (-1)^{i_x+i_y} h \hat{S}_i^z$
where $\hat{S}_i^z = (\hat{n}_{\uparrow i}-\hat{n}_{\downarrow i})/2$ to induce a local antiferromagnetic order, where $h$ controls the strength of pinning and is chosen as 
$h=1/2$ unless otherwise specified.
When pinning is present, energies are reported including the contribution from the additional field, consistent with other publications \cite{Xu2022}. 

Our implementation utilizes Jax \cite{jax2018github} to perform automatic differentiation and sampling, alongside the Optax package \cite{deepmind2020jax} for optimization. The matrix exponential is computed approximately using a $k=3$ product formula \cite{Bader2019}. Optimization generally uses around 180 optimization steps with varying and decaying step size (see appendix); gradients are accumulated with 189,000 samples and parameters are updated using a random step with the computed sign \cite{Luo2019}. 

This work focuses on the case of $N_\ell=4$, which gives substantial improvement over e.g. $N_\ell=2$ or a single slice. While any given calculation can always be re-optimized with additional slices, by fixing $N_\ell$ we can observe the limits of this form. 

\begin{figure}[th]
    \centering
    \includegraphics[width=\columnwidth]{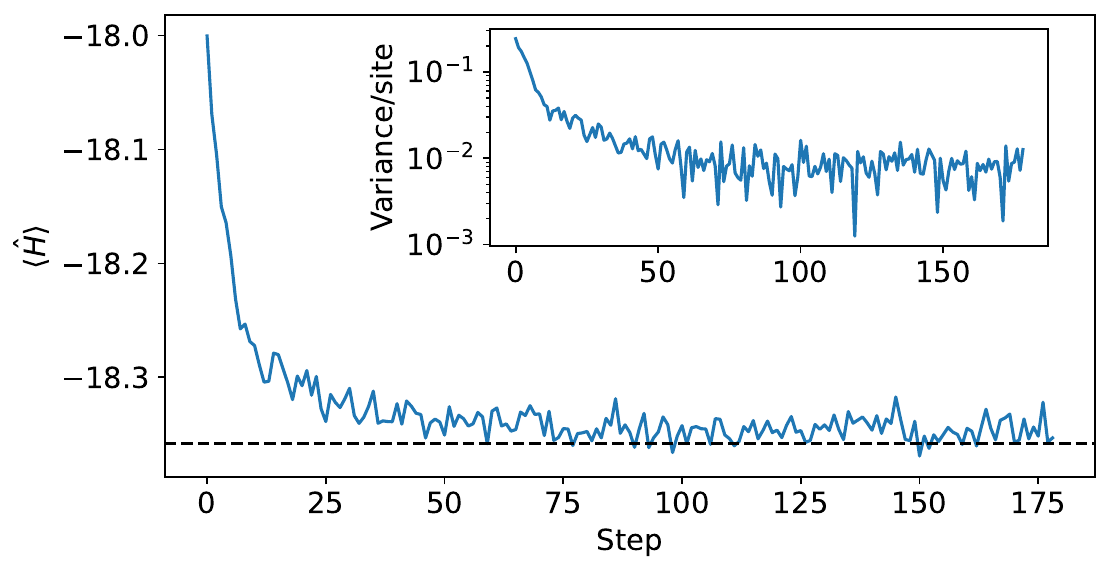}
    \caption{Variational energy $\braket{\hat{H}}$ and variance per site $(\braket{\hat{H}^2}-\braket{\hat{H}}^2)/N$ (inset) during optimization of a $4\times 4$ Hubbard model at $U=6$ and $n=0.625$ using VAFQMC with $N_\ell = 4$. The ground state from exact diagonalization is shown in a black dashed line. Fluctuations are due to the limited sampling during optimization. }
    \label{fig:4x4_energy_opt}
\end{figure}

We can additionally add a term to optimize the sign similar to Ref.~\cite{chenHAFQMC}, however in the applications presented here the average sign is almost always very large $\braket{S} \geq 0.93$. For $U=8$, the sign can get as low as $\braket{S} \approx 0.74$ but this is sufficiently large %
to resolve observables and derivatives without needing any further
constraint in the optimization. 
We renormalize (i.e. `clip') both the sign term derivative and the energy derivative term to be on the same order. An example of the optimization is shown in Fig.~\ref{fig:4x4_energy_opt}. 

After optimization, parameters are selected based on the lowest observed average energy. Using this parameter set, observables such as the local energy and electron occupation are measured with approximately $10^6$ samples with a thermalization of $10^3$ sweeps. In order to properly treat the $1/\langle S\rangle$ term, error bars are reported from a jackknife estimate \cite{becca2017quantum}. 

DMRG results were computed using the ITensor library \cite{ITensor,ITensor-r0.3} with a maximum bond dimension of $m=15360$. 
Constrained-path AFQMC results are obtained using protocols similar to those 
in Refs.~\cite{Xu2023, Xu2022,Qin2020,Vitali2019} including self-consistently optimized constraints \cite{Qin2016}.

\section{Results}\label{sec:results}

\subsection{Energetic Benchmarks}

\begin{table*}
    \centering
    \resizebox{2\columnwidth}{!}{
\begin{tabular}{rl|r@{\extracolsep{0pt}.}l|r@{\extracolsep{0pt}.}lr@{\extracolsep{0pt}.}lr@{\extracolsep{0pt}.}lr@{\extracolsep{0pt}.}lr@{\extracolsep{0pt}.}lr@{\extracolsep{0pt}.}l|r@{\extracolsep{0pt}.}lr@{\extracolsep{0pt}.}lr@{\extracolsep{0pt}.}l}
$N_{x}\times N_{y}$  &  & \multicolumn{2}{c|}{$6\times6$} & \multicolumn{2}{c}{} & \multicolumn{10}{c|}{$8\times8$} & \multicolumn{2}{c}{} & \multicolumn{4}{c}{$10\times10$}\tabularnewline
$n$  &  & \multicolumn{2}{c|}{$2/3$} & \multicolumn{2}{c}{} & \multicolumn{2}{c|}{$0.6875$} & \multicolumn{4}{c|}{$0.78125$} & \multicolumn{4}{c|}{$0.875$} & \multicolumn{2}{c}{} & \multicolumn{4}{c}{$0.8$}\tabularnewline
$U$  &  & \multicolumn{2}{c|}{4} & \multicolumn{2}{c}{} & \multicolumn{2}{c|}{4} & \multicolumn{2}{c}{4} & \multicolumn{2}{c|}{8} & \multicolumn{2}{c}{4} & \multicolumn{2}{c|}{8} & \multicolumn{2}{c}{} & \multicolumn{2}{c}{$4$} & \multicolumn{2}{c}{$8$}\tabularnewline
\hline 
\hline 
Projector Methods  &  & \multicolumn{2}{c|}{} & \multicolumn{2}{c}{} & \multicolumn{2}{c}{} & \multicolumn{2}{c}{} & \multicolumn{2}{c}{} & \multicolumn{2}{c}{} & \multicolumn{2}{c|}{} & \multicolumn{2}{c}{} & \multicolumn{2}{c}{} & \multicolumn{2}{c}{}\tabularnewline
$E_{AFQMC}/N$  &  & -1&18525(4){*}  & \multicolumn{2}{c}{} & -1&1858(2){*}  & -1&13253(3) & -0&9252(1) & -1&01923(6)  & -0&7616(1)  & \multicolumn{2}{c}{} & -1&1135(2){*}  & -0&8960(1) \tabularnewline
$E_{FN}/N$  &  & \multicolumn{2}{c|}{} & \multicolumn{2}{c}{} & \multicolumn{2}{c}{} & -1&128396(9) & -0&913753(9) & -1&01153(1)  & -0&74941(2)  & \multicolumn{2}{c}{} & -1&10934(1)  & -0&88291(1)\tabularnewline
\hline 
Variational Methods  &  & \multicolumn{2}{c|}{} & \multicolumn{2}{c}{} & \multicolumn{2}{c}{} & \multicolumn{2}{c}{} & \multicolumn{2}{c}{} & \multicolumn{2}{c}{} & \multicolumn{2}{c|}{} & \multicolumn{2}{c}{} & \multicolumn{2}{c}{} & \multicolumn{2}{c}{}\tabularnewline
$E_{VMC}/N$  &  & \multicolumn{2}{c|}{} & \multicolumn{2}{c}{} & \multicolumn{2}{c}{} & -1&12683(1) / & -0&91011(2) / & -1&00579(1)  & -0&74220(3)  & \multicolumn{2}{c}{} & \multicolumn{2}{c}{} & \multicolumn{2}{c}{}\tabularnewline
 &  & \multicolumn{2}{c|}{} & \multicolumn{2}{c}{} & \multicolumn{2}{c}{} & -1&13037(1)$^{\dagger}$ & -0&91482(1)$^{\dagger}$ & \multicolumn{2}{c}{} & \multicolumn{2}{c|}{} & \multicolumn{2}{c}{} & \multicolumn{2}{c}{} & \multicolumn{2}{c}{}\tabularnewline
$E_{VAFQMC}/N$  &  & -1&1834(1)  & \multicolumn{2}{c}{} & -1&18419(7)  & -1&13048(8) & -0&9148(1) & -1&01391(8)  & -0&7469(2)  & \multicolumn{2}{c}{} & -1&11097(6)  & -0&8819(1)\tabularnewline
\hline 
\hline 
$\sigma_{VMC}^{2}$  &  & \multicolumn{2}{c|}{} & \multicolumn{2}{c}{} & \multicolumn{2}{c}{} & 1&43 / 0.59(2)$^{\dagger}$ & 3&46 / 2.58(1)$^{\dagger}$ & 2&79  & 6&62  & \multicolumn{2}{c}{} & \multicolumn{2}{c}{} & \multicolumn{2}{c}{}\tabularnewline
$\sigma_{VAFQMC}^{2}$  &  & 0&2(1)  & \multicolumn{2}{c}{} & 0&3(3)  & 0&5(3) & 2&4(4) & 0&6(3)  & 3&5(5)  & \multicolumn{2}{c}{} & 0&8(6)  & 3&9(8)\tabularnewline
\end{tabular}
}
    \caption{Energy comparison between AFQMC \cite{LeBlanc2015,Shi2013}, 
    fixed-node DMC \cite{LeBlanc2015}, 
    VMC \cite{varbench}, and VAFQMC (this work), on a PBC square Hubbard model with  density
    $n$ and interaction strength $U$ (in units of $t$). Energies with a * are effectively exact from release constraint calculations of AFQMC \cite{Shi2013}, and energies with $\dagger$ include SU(2) and momentum projection augmented with a Lanczos step \cite{varbench}. Additionally shown is the variance $\sigma^2 = \braket{H^2}-\braket{H}^2$ of the energy, which can be computed more conveniently in  variational methods and which
    vanishes  ($\sigma^2 = 0$)
    for a true eigenstate of the Hamiltonian. VAFQMC results are optimized with $N_\ell=4$ projector slices. 
    }
    \label{tab:square_energy_bench}
\end{table*}

To test the variational ansatz, we first benchmark against doped large square Hubbard models with PBC as shown in Table~\ref{tab:square_energy_bench}. Some systems at intermediate doping have a reference value from AFQMC \cite{Shi2013} which is often effectively exact; we compare to AFQMC the following systems: $6\times 6$ with $n=2/3$, $8\times 8$  with $n=0.6875$, and $10\times 10$ with $n=0.8$ at $U/t=4$.  The VAFQMC ansatz is able to obtain an energy error that is $O(10^{-3})$ with an energy variance per site $(\braket{H^2}-\braket{H}^2)/N$ of less than 1, 
which improves over traditional variational
approaches in such systems.

For systems without exact energies, we compare with several 
state of the art methods, projector and also variational QMC methods. 
Note that of the projector methods, AFQMC and fixed-node projector 
diffusion Monte Carlo (DMC) 
with a VMC optimized trial wave function, only fixed-node is variational. Then we compare with VMC energy and energy variance. Additionally note that the fixed-node DMC energies of Ref.~\cite{varbench} are higher than those of Ref.~\cite{LeBlanc2015}; we report the lowest energy available.

Overall, the variational energy of the VAFQC ansatz is competitive or improves upon those from state-of-the-art VMC under the standard formalism of Slater determinants (or 
antisymmetrized germinal powers), backflow, and Jastrow factors. 
For the $8\times 8$ periodic system, at $U=4$, VAFQMC is lower in energy across densities and, 
at $U=8$, $E/N$ from VAFQMC is approximately $0.004t$ lower at $n=0.78125$ and 
$0.01t$ higher at $n=0.875$. 
More impressively, at $n=0.78125$ we obtain nearly the same energies for $U=4$ and $U=8$ (with $U=8$ having worse performance) to a very sophisticated VMC wave function, enforcing symmetries such as SU(2) and $k=0$ momentum projection, augmented by additional Lanczos steps.
Likewise when comparing to fixed-node DMC energies across system sizes and 
densities, we find similar or slightly lower energies for $U=4$ and competitive energies for $U=8$. This is particularly favorable as VAFQMC can conveniently compute
both local and non-local observables and correlation functions,
which can be challenging or inaccessible within fixed-node DMC and which require back-propagation \cite{Zhang1997} in AFQMC. 

We also note for the e.g. $10\times 10$ $U=4$ point, the local magnetization $\langle S^z\rangle$ is $<10^{-2}$ on each site, recovering an unpolarized spin solution despite the initial parameters breaking spin symmetry explicitly. This result comes at a small cost however, and lower energy results can be obtained when starting from a UHF trial wave function (e.g. Fig.~\ref{fig:cartoon}c). Using a UHF initial state $\ket{\phi_0}$ with $U_{UHF}=0.5$ (the same as used by AFQMC for its trial wave function), the local magnetization is $\braket{S^z_i} \sim O(10^{-2})$ but about $0.05t$ lower in energy. This is not unique to VAFQMC but rather to variational wave functions generally, where local minima of symmetry broken states can be hard to optimize away. In VAFQMC this manifests as requiring more projector slices $N_\ell$ to modify a strong local magnetization or density in the trial state, compared to a spin-restricted initial ansatz, and thus we suggest utilizing the restricted form to minimize total parameters. 
 
\subsection{Local Order Prediction
and Automatic Symmetry Restoration}

\begin{figure*}[ht]
    \centering
  \includegraphics[width=2\columnwidth,trim={0 7pt 0 7pt},clip]{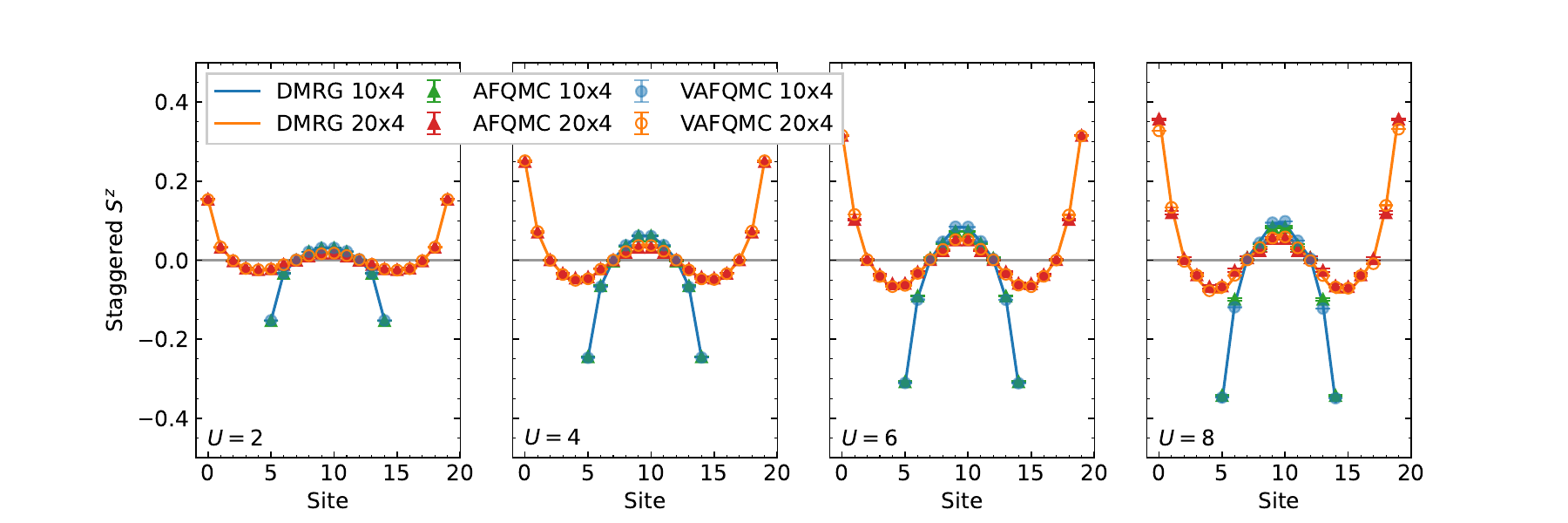} \\
  \includegraphics[width=2\columnwidth,trim={0 7pt 0 7pt},clip]{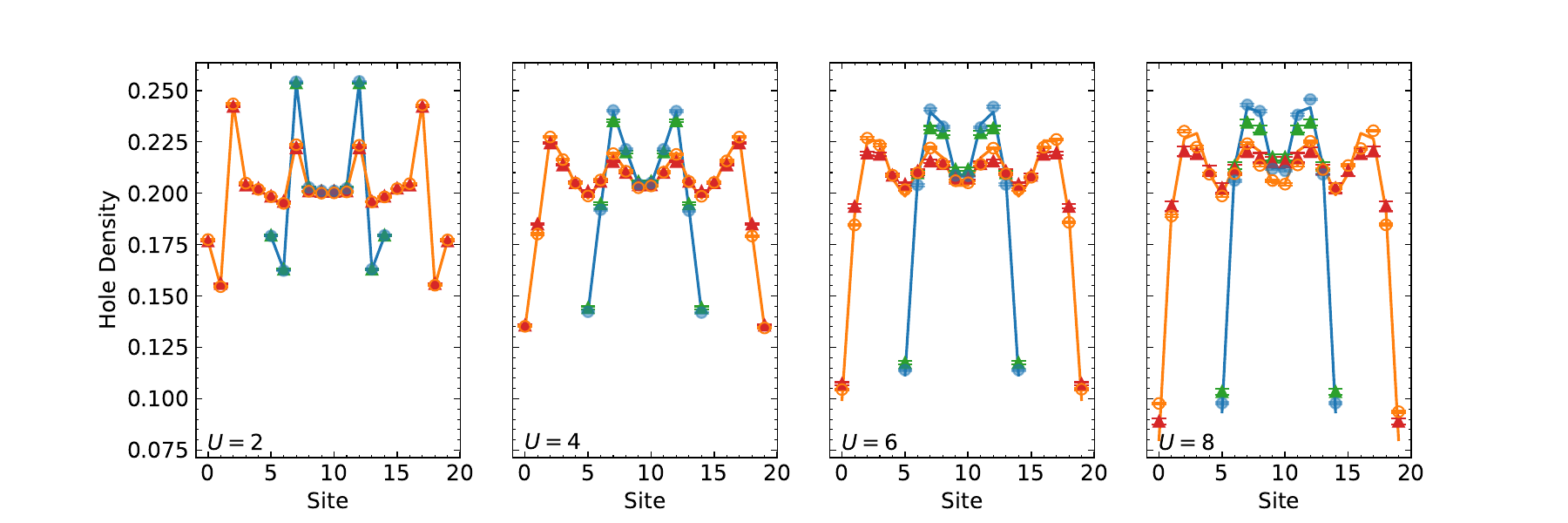}
    \caption{Comparison of staggered $S^z$ $\langle (-1)^{i_x+i_y}S^z(i_x,i_y)\rangle $ (top) and hole density $\langle 1-n(i_x,i_y)\rangle $ (bottom) between DMRG, VAFQMC, and AFQMC for a Hubbard model on a cylinder with edge pinning at a series of interaction strengths $U=2,4,6,8$ and $n=0.8$ density. Results have been averaged over the width-4 column direction. VAFQMC results are with $N_\ell=4$ projector slices. 
    \label{fig:cyl_local_ord}
    }
    
\end{figure*}

\begin{figure*}[t]
    \centering
    \includegraphics[width=2\columnwidth,trim={0 7pt 0 0},clip]{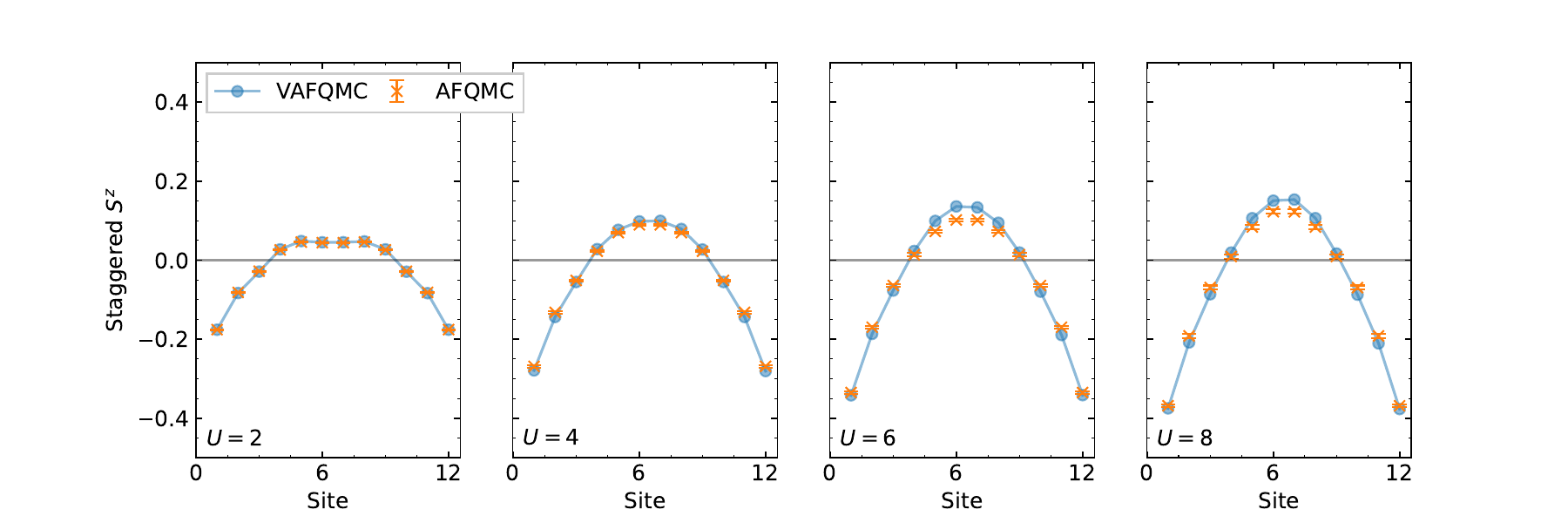}\\
       \includegraphics[width=2\columnwidth,trim={0 0 0 7pt},clip]{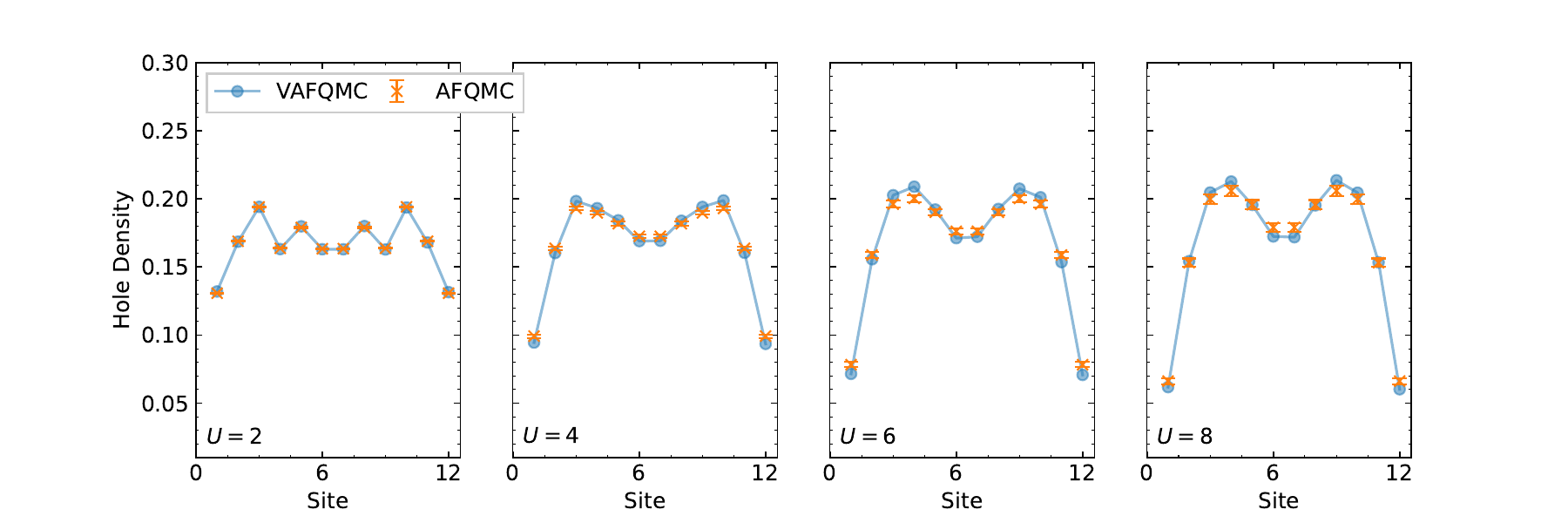}
    \caption{Comparison of staggered $S^z$ $ \langle (-1)^{i_x+i_y}S^z(i_x,i_y)\rangle $ (top) and hole density $\langle 1-n(i_x,i_y)\rangle $ (bottom) between VAFQMC and AFQMC for a $12\times 8$ Hubbard model on a cylinder with edge pinning at a series of interaction strengths $U=2,4,6,8$ and $n=5/6$ density. Results have been averaged over the width-8 column direction. VAFQMC results are with $N_\ell=4$ projector slices. }
    \label{fig:cyl_x8}
\end{figure*}

As a proxy for phase diagram exploration, we use VAFQMC to explore the local order of cylindrical Hubbard models with antiferromagnetic edge pinning. This model was used to study the stripe behavior of the ground state \cite{Zheng2017,Xu2022},
In Ref.~\cite{Xu2022}, the stability of the stripe state was determined by comparing the staggered spin and hole densities at different interaction strength $U$ and electron density $n$ as a function of system size. We perform calculations similar to those, but first with small enough system sizes that DMRG can provide quasi-exact reference results. 
We vary the
electron density $n$, or equivalently hole 
doping of $\delta\equiv 1-n$, and 
consider cylinders %
of width $r/\delta \times N_y$ for some integer $r$. 

With different interaction strengths and system sizes, we find remarkable agreement in the magnetic order between VAFQMC, DMRG, and AFQMC in the top of Fig.~\ref{fig:cyl_local_ord}. The variational ansatz for VAFQMC is completely unconstrained by symmetries, yet can nearly recover both translational symmetry in the $y$ direction and symmetry across the middle of the cylinder. 
The corresponding  hole densities are shown in bottom of Fig.~\ref{fig:cyl_local_ord}.
We again see almost complete recovery of the symmetry of the system. 
There is small disagreement between DMRG and AFQMC as to the hole densities  
at large $U$. As DMRG is expected to be nearly exact in these width-4 systems,
the discrepancy is likely an indication of small residual bias from the 
constraint in AFQMC \cite{Qin2016}.
Remarkably our VAFQMC results are nearly indistinguishable from the DMRG results. 
This extends to other dopings across similar values of $U$ as well (see appendix),
and  illustrates the predictive power of the VAFQMC method.  
At larger values of $U$, the variational energy from VAFQMC is 1-2\% higher than DMRG and AFQMC. These results are a reminder that the total energy, while 
very important for a variational ansatz, should not be unduly emphasized, and 
order parameters and correlation functions often provide a more stringent measure of the predictive power of an ansatz. 

We next move beyond width-4 cylinders and
consider wider systems, which are
important \cite{Xu2022,Xu2023} to access the thermodynamic limit. 
In Fig.~\ref{fig:cyl_x8} a comparison of staggered $S^z$ and hole density between AFQMC and VAFQMC is shown for a $12\times 8$ cylinder with $n=5/6$ density and $U=2,4,6,8$. Exceptionally, VAFQMC continues to obtain the correct order and stripe wavelength, with reasonable quantitative agreement with AFQMC. 
Combined the results for larger periodic cells discussed in the previous 
section, these results show the promise of the VAFQMC ansatz 
as a predictive approach for extended two-dimensional systems and beyond.

\begin{figure*}[th]
    \centering
    \includegraphics[width=\columnwidth]{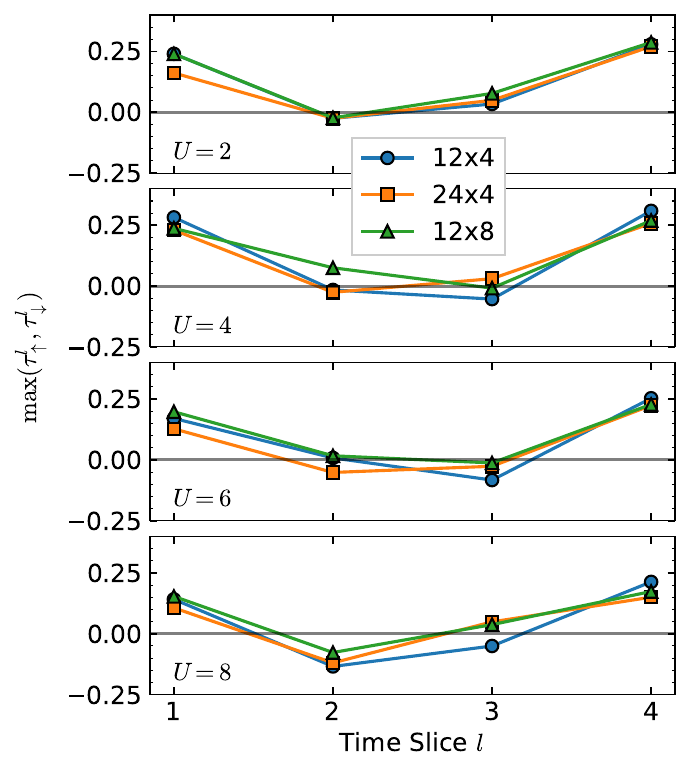}
    \includegraphics[width=\columnwidth]{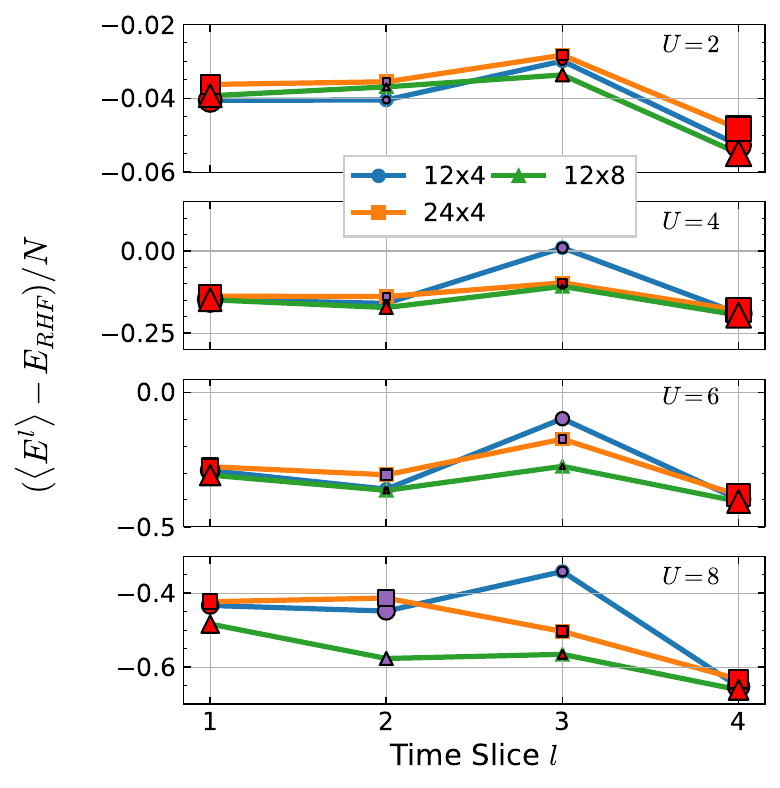}
        
    \caption{Left: Optimized values for $\tau^l$ maximized over spin ($\tau^l = \max(\tau^l_\uparrow, \tau^l_\downarrow)$) %
    for the Hubbard model in cylindrical geometry with edge pinning,
    shown for four values of $U$ and three system sizes.
    Values have been renormalized by $\langle T^{\sigma l}_{\langle ij\rangle} \rangle $. The electron density is $n=5/6$. 
    Right: Energy %
    per site (relative to that from $|\psi_{RHF}\rangle = |\phi_0\rangle$) at a given time slice $l$ (in units of $t$) for the systems shown on the left.
    Note that the energy at each $l$ is 
    evaluated with the final
    parameters fixed at the 
    optimized values for $l=4$.
    Marker size is proportional to the renormalized $\tau^l = \max(\tau^l_\uparrow, \tau^l_\downarrow)$ with red denoting positive (equivalent to cooling projection) and purple denoting negative (equivalent to heating projection). Error bars are smaller than the marker size. } 
    
    \label{fig:cyl_times}
\end{figure*}

As mentioned, the projection times are unconstrained as full variational parameters in our approach.
Unlike the ansatz in Ref.~\cite{sorellaVAFQMC}, a number of solutions have a negative $\tau^l$ value within one of the projector slices,
as shown in the left of Fig.~\ref{fig:cyl_times},
similar to that of higher order Trotter expansions \cite{Ostmeyer2023,Janke1992}. As both $T^{\sigma l}_{ij}$ and $\tau^l$ can change, we renormalize the $\tau^l$ values by the average value of nearest-neighbor hopping terms $\langle T^{\sigma l}_{\langle ij\rangle}\rangle$ which is almost always $O(-t)$.  
By only considering $\tau^l$ values, we can see that the effect of each time slice is not equivalent over a given $U$. As $U$ increases, the value of $\tau^l$ tends to fluctuate between positive and negative.
We believe this is to effectively `raise the temperature' and  broaden the trial wave function in order to (in the next slice) project into the correct manifold. 
For width-4 systems, $|\tau^l|$ values are around $\approx 0.13$ while width-8 systems they are slightly larger at $0.15$ on average, but both width cylinders have slices $l$ where $\tau^l_\sigma \approx 0.25$ for example. Width-4 systems also tend to have larger negative $\tau^l_\sigma$, with $\min_{l,\sigma} \tau^l_\sigma \sim -0.2$ for width-4 cylinders and $\min_{l,\sigma} \tau^l_\sigma \sim -0.07$ for width-8. We emphasize that while some values of $|\tau^l|$ are quite small, the interaction term remains $O(10^{-1})$ and thus the ansatz is still not reduced to fewer slices.

To examine %
the role that the negative $\tau$ values play, in the right of Fig.~\ref{fig:cyl_times} we plot the $\langle E^l\rangle $ for the optimized four slice VAFQMC ansatz truncated to $l$
($l\leq 4$) time slices. (This is to be distinguished from a stand-alone $l$-slice or $N_\ell =l$ wave function.
In the latter the parameters would be optimized for $l$ slices, while here they are 
optimized for the entire 4-slice wave function.)
We subtract the energy of that of the $\ket{\phi_0}$ state (denoted \textit{RHF}) and rescale by $1/N$. We see that for the first ($l=1$) and last ($l=4$) slice the energy always decreases. At intermediate values of $l$ there may be energy values that are significantly worse than the previous time slice, unlike the near monotonic decrease in %
a QMC projector method. For example consider the $12\times 4$ system at $U=4$, the energy per site at an intermediate point $l=2\to 3$ jumps by $\Delta E/N \sim +0.17$, 
then lowers in the final slice $(l=3\to4)$ %
by $\Delta E/N \sim -0.2$.
The large value of $|\tau^4|$ at the end of the ansatz appears to be linked to this substantial decrease in energy, to compensate for the heating and cooling of the path.

For lower values of $U$, often a given slice may have a very small value of $\tau^l$, suggesting that an ansatz with fewer $N_\ell$ may suffice to capture the dominate physics. This 
is further suggested by the intermediate values of the energy $\braket{E^l}$, which only occupies a range of $\sim 0.04$/site.
We conjecture that the flexibility in $\{ \tau^l\}$ can improve the size extensivity of the ansatz. 
The ansatz can both increase the overall effective $\beta$, similar to Ref.~\cite{sorellaVAFQMC}, to produce a size extensive ansatz, or may be something more complex within the path.
The new ansatz based on an effective Hamiltonian
given in Eq.~(\ref{eq:effectiveH}) is the key 
for allowing better size consistency with a 
fixed  $N_\ell$.
More systematic studies on this will be very valuable.

\section{Discussion and Summary}\label{sec:conclusion}

We presented a variational approach for many-electron systems with an expressive ansatz of $\mathcal{O}(N_\ell\cdot N^2)$ parameters. 
Although inspired by the idea of an approximate realization of the imaginary-time projection in AFQMC, the ansatz is much more general. 
The approach shares the idea of variational optimization 
of a small number of projections with AFQMC-like time slices as in
Ref.~\cite{sorellaVAFQMC}, and we have retained
the name VAFQMC from there. 
However, our approach takes a 
different philosophy of viewing the ansatz as a
variational {\it effective\/} Hamiltonian rather than 
a variation projection using the {\it physical\/} Hamiltonian. 
Instead of retaining the form of the Hamiltonian $\hat H$ in the projection
(e.g., keeping $\hat T$ and the basic form of the 
HS transformation dictated by $\hat V$) and 
creating different orders via $|\phi_0\rangle$,
we view the ansatz as seeking the most effective projection (to the true ground state) via variational optimization.

We have emphasized the underlying theory of optimizing 
a set of effective one-body Hamiltonians or actions defined by auxiliary fields. 
This wave function, even in our still rudimentary implementation, 
is already competitive with state of the art VMC and even fixed-node Green's function Monte Carlo. We have demonstrated this on
large square Hubbard models, where the approach can systematically obtain 
the ground state energy to within or better than $1$-$2\%$ of the exact results.
Furthermore, it nearly restores symmetry automatically via its optimization process and
provides 
accurate physical observables which allow predictive resolution of magnetic and charge 
correlations and orders.
Broken e.g. $SU(2)$ symmetry from pinning fields can be additionally restored by hand and used in the original model. 

Historically, many VMC wave function ansatze were built around a specific mean-field state (for example in  Ref.~\cite{sorellaVAFQMC}), or 
an explicit form that targets a particular type of order.
Under this approach, optimization is constrained to the fixed state type,
and prediction of the actual order is made by 
comparing the relative energetic behaviors of two or more such fixed types of states.
Despite its considerable success, 
this approach lacks true predictive power
in many situations, either because the 
order is not known, or because the different types being compared are not 
well balanced in the ansatz (e.g., one with more variational flexibility than the other). With the onset of neural quantum states (e.g. Refs.~\cite{Carleo2017,Luo2019,RobledoMoreno2022}), wave functions are parameterized with partially or fully unconstrained variational freedom to improve expressibility. 
Likewise, the aim of our VAFQMC approach is to have an ansatz which is similarly more flexible and more expressive such that
local order can instead be predicted from optimization. 
As our examples have shown, with large optimization freedom 
VAFQMC can function as a \textit{predictive} ansatz in addition to providing good variational energies in a challenging system operating at  realistic parameters and under ``real-life'' conditions. It is also notable that this is achieved without the use of neural networks or tensor networks, bringing a new class of unconstrained systemically improvable variational wave functions. 

Optimization of an expressive ansatz alone is sometimes
still not enough to
resolve 
the hardest problems. For models in which there are many competitive states, e.g. stripes or intertwined orders, the application of 
pinning fields has become a very useful and 
essential tool
to elucidate the underlying order. The success with which this paradigm has been used \cite{Assaad2013,Zheng2017,Qin2020,Xu2022,Xu2023},  
and this benchmark study of VAFQMC suggests that this is a potential paradigm to better solve problems inaccessible to 1D tensor networks. This should prove to be an invaluable method in the toolbox of computational many-body physics.   

Despite the `inexpensive' ansatz presented here with fixed $N_\ell$, we emphasize 
that VAFQMC can still be systematically improved. One can continue to add more projector slices to potentially improve the energy, but after some point computational costs will increase
(and further care must be taken to properly stabilize the `long time' projection of the trial state). At $N_\ell=4$ this was functionally unnecessary for the systems we considered. 
Neural networks can be added within the ansatz,
for example as in Ref.~\cite{chenHAFQMC},
to further improve expressibility, particularly in the strongly correlated regime where correlations between auxiliary fields may be useful. 
Additional considerations may be needed to 
improve scalability, however.
Likewise while the optimization procedure was able to regain translation invariance and minimal local spin symmetry breaking, this should not be relied upon when scaling. Applying symmetry projection \cite{Shi2014,Sheikh2021}, either within $\ket{\phi_0}$ or for each projected $\ket{x}$ state, is a compelling future direction. 

Given the promise the ansatz has already shown at this 
very early stage of development, we believe the method presents 
many exciting and interesting new possibilities. 
In addition to the directions discussed above, there are  
a number of issues  
to be addressed or better understood.
For example, optimizing to machine precision is currently intractable 
due to an infinite variance problem. Because the local energies do not have a zero variance principle, sampling can be dominated by rare events and thus are unable to reduce the statistical noise with more samples. We have checked our results on small systems using the methods of Refs.~\cite{Shi2016} (see appendix), however there may be a better methods to be explored in future work.

\begin{acknowledgments}
We would like to thank Di Luo and Yixiao Chen for useful discussions, and 
Yiqi Yang for providing AFQMC results. The Flatiron Institute is a division of the Simons Foundation. 
\end{acknowledgments}
\bibliography{main}

\appendix
\renewcommand{\thefigure}{A\arabic{figure}}

\section{Further Computational Details}
The step sizes during the first 79 steps of optimization is 0.01 and then is lowered to 0.001 for the remaining 100 steps. The optimal point is selected by the lowest energy during the optimization, normally somewhere after the first 80 steps. 

Optimization was performed with 126 independent walkers contributing to the total sample collection, while the additional measurement of the optimized parameters used 120 walkers. Each measurement was collected after $N\cdot N_\ell$ Metropolis updates and values reported in the main text were averaged over walkers and a 1000 sweep thermalization time was used. 

Finally the initial values were chosen to correspond to a $\tau_\ell=0.1$ projection of AFQMC, with a perturbation of $\delta\in [0,10^{-3})$ for $\lambda$ and $T^\ell_{\sigma}$ terms. Note that when there is edge pinning, we do not include pinning terms in $T^\ell_{\sigma}$ and $\ket{\phi_0}$ but do include the appropriate cylindrical boundary conditions.  Variance measurement was performed with an additional 10x samples.  

\section{Additional Hubbard Cylinder Comparison}

\begin{figure*}
    \centering
    \includegraphics[width=2\columnwidth]{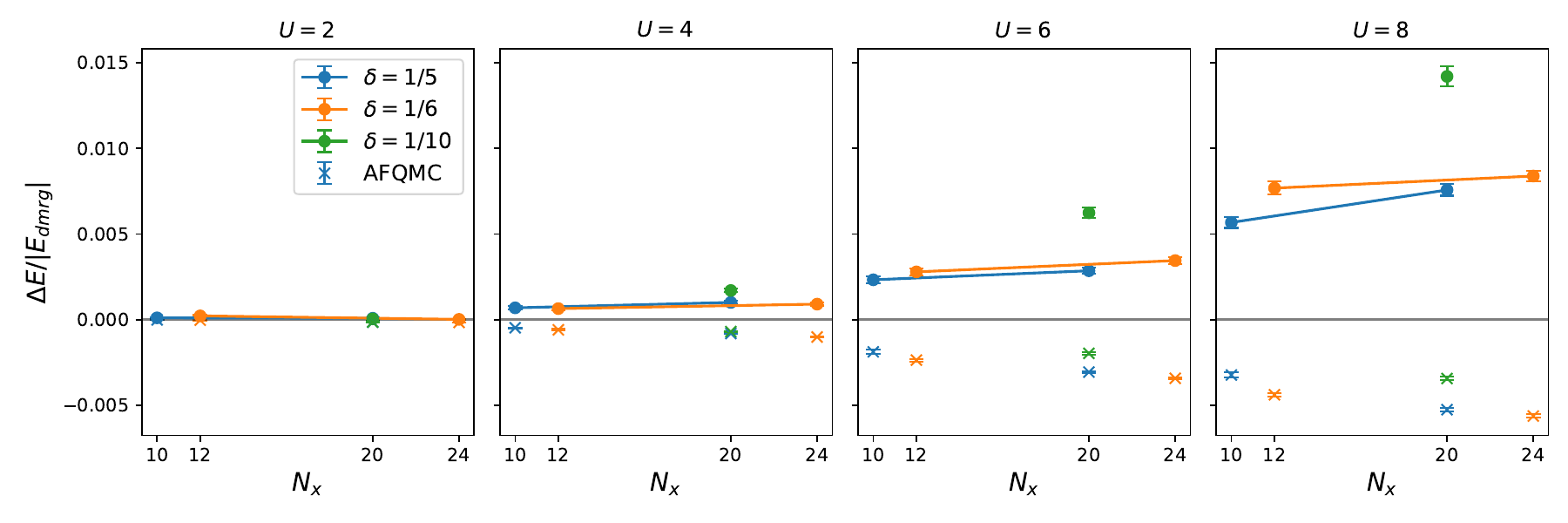}
    \caption{Relative energy $(\langle E \rangle - E_{DMRG})/|E_{DMRG}|$ comparison for VAFQMC and CP-AFQMC compared to DMRG for various Hubbard models of size $N_x \times 4$ on a cylinder with edge pinning and hole doping $\delta$.   }
    \label{fig:app_cyl_energy}
\end{figure*}

\begin{figure*}
    \centering

    \includegraphics[width=2\columnwidth,trim={0 7pt 0 0},clip]{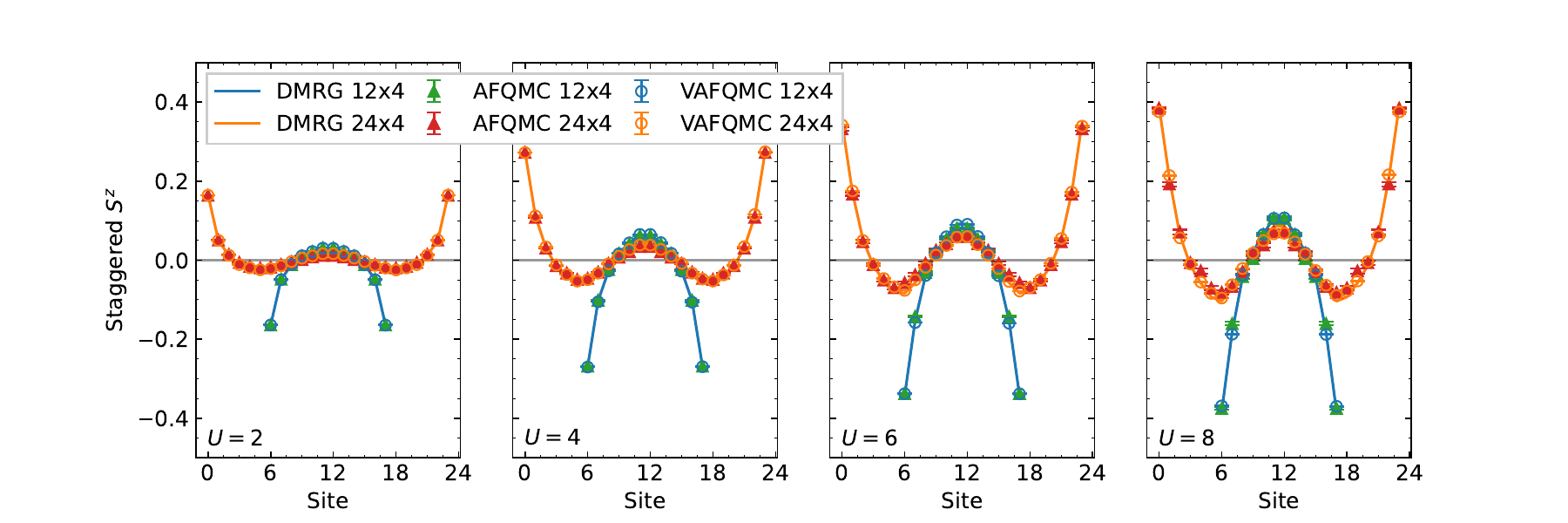} \\
    \includegraphics[width=2\columnwidth,trim={0 0 0 7pt},clip]{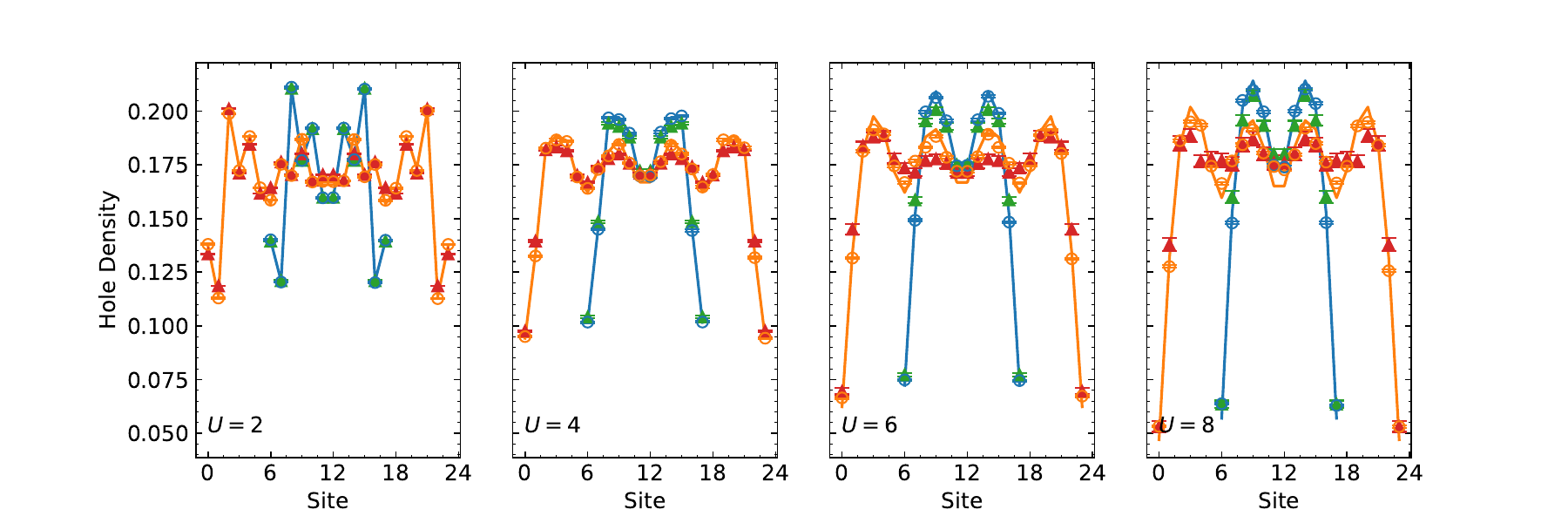}  
    
    \caption{Comparison of staggered $S^z$ $\langle (-1)^{i_x+i_y}S^z(i_x,i_y)\rangle $ (top) and hole density $\langle 1-n(i_x,i_y)\rangle $ (bottom) between DMRG, VAFQMC, and AFQMC for a Hubbard model on a cylinder with edge pinning at a series of interaction strengths $U=2,4,6,8$ and $n=5/6$ density. Results have been averaged over the width-4 column direction. VAFQMC results are with $N_\ell=4$ projector slices.    }
    \label{fig:app_cyl_order_sixth}
    
\end{figure*}

We provide both additional data from the cylindrical Hubbard model systems with edge pinning studied above along with data for $\delta= 1/10$ which is shown in Ref.~\cite{Xu2022}. This data is again compared to those from 
DMRG and self-consistent CP-AFQMC
in width-4 cylindrical systems.

In Fig.~\ref{fig:app_cyl_energy}, the energy of VAFQMC and AFQMC is compared to that of the obtained DMRG. Note that the DMRG results are not extrapolated in cutoff, and thus may be slightly higher in energy to the exact ground state energy, which is unknown. At $U=2$, all methods are in agreement for the the energy within $10^{-4}$ relative difference. For larger values of $U$, AFQMC  results trend lower in energy as $U$ gets larger while VAFQMC results trend higher. The largest discrepancies occur at $U=8$ where VAFQMC is approximately $1-1.5\%$ from DMRG while AFQMC is $<1\%$ lower (relatively). 

We show further local observables comparisons, of average staggered $S^z$ and hole density, at $n=5/6$ for $12\times 4$ and $24\times 4$ system sizes in Fig.~\ref{fig:app_cyl_order_sixth}. The VAFQMC results of spin and hole densities mirror those at in Fig.~\ref{fig:cyl_local_ord} 
are seen to be in excellent agreement with those from DMRG. While staggered spin densities agree well between the three methods, VAFQMC and DMRG agree on the magnetic order, the charge order has slight discrepancies where VAFQMC and DMRG more closely agree than with AFQMC.

\section{Infinite Variance Problem}
The estimator of the variational energy $\langle E \rangle = \braket{E_L S} / \braket{S}$ can suffer from an infinite variance problem~\cite{Shi2016}, as  $\braket{x^\prime|x}$ approaches zero. In order to consider the effect on our results, we use the `bridge link' mitigation of Ref.~\cite{Shi2016}. A new sampling probability distribution is introduced, by inserting an additional (symmetric) time slice in the center of the path, i.e.
\begin{equation}
    \rho^\prime(x,x^\prime) = |\braket{x^\prime|e^{-\tau H}|x}|.
\end{equation}
The projector introduces an additional set of $N$ auxiliary fields to integrate over, and modifies the energy calculation to be 
\begin{align}
    \braket{E} &= \frac{\braket{\psi|\hat{H}|\psi}}{\braket{\psi|\psi}} \\
    &= \frac{ \braket{E_L \, \rho/F}_{\rho^\prime }}{\braket{ \rho/F}_{\rho^\prime}}\label{eq:energy_inf}
\end{align}
where $F=F(\tau) = |\braket{x^\prime|e^{-\tau \hat{H}}|x}|$ and $\rho = \braket{x^\prime|x}$ which has a sign unlike previously. The calculation of $F$ must be done with care however, so as to not introduce a estimator bias (see \cite{Alexandru2023}).

\begin{figure}[t]
    \centering
    \includegraphics[width=\columnwidth]{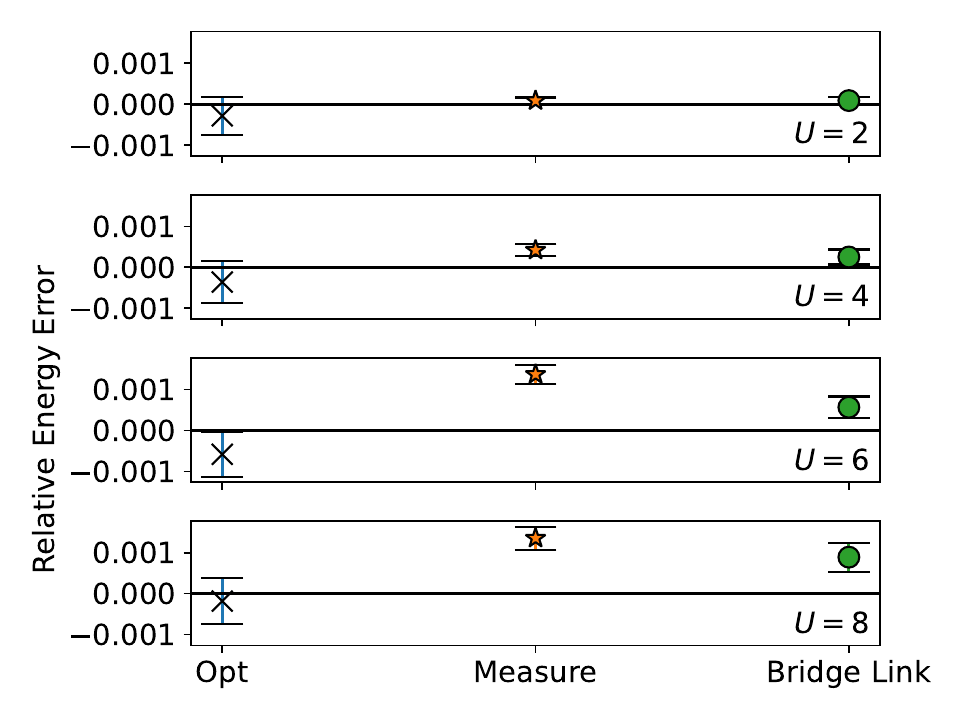}
    \caption{Relative energy error $(\braket{E}-E_0)/|E_0|$ between VAFQMC and the exact $E_0$ from exact diagonalization of a $4\times 4$ Hubbard model with PBC at $n=0.625$. Blue `x's (``Opt"), are the energies reported from optimization, orange stars (``Measure") are measured with methods equivalent to sec.~\ref{sec:methods}, and green circles (``Bridge Link") represent bridge link calculations where $F(\tau=0.005)$ is summed exactly . }
    \label{fig:energy_comparison}
\end{figure}

In order to consider the effects of the infinite variance, we study a closed-shell configuration ($n=0.625$) of the $4 \times 4$ Hubbard model with PBC at various values of $U$. The calculation of $F$ is done exactly with a bridge link of $\tau=0.005$, summing the $2^{16}$ configurations for each field configuration of $x,x^\prime$.  In Fig.~\ref{fig:energy_comparison} we show the energy for both the standard estimator of Eq.~(\ref{eq:energy}) during and after VAFQMC optimization, and with included bridge link Eq.~(\ref{eq:energy_inf}). Optimized energies are often below the exact energy in part due to small sample sizes. These are then raised (potentially above the ground state) with increased measurement, and subsequently the inclusion of the bridge link shifts the energy slightly slower but still within error bars of the original measurement estimate.

\section{Additional $\tau^l$ data}
\begin{figure}[t]
    \centering
    \includegraphics[width=\columnwidth]{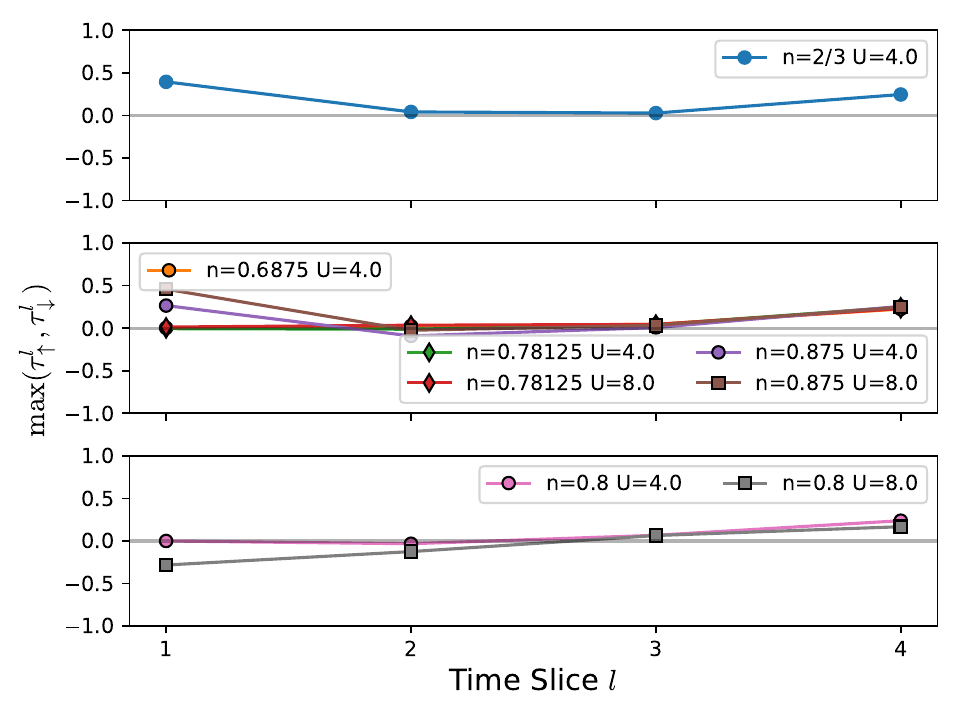}
    \caption{Optimized values for the maximum $\tau^l$ (maximized over spin)  for 2D Hubbard models of size (top to bottom) $6\times 6$, $8\times 8$ and $10\times 10$ with PBC at $U$ and $n$ values as shown.  Values have been renormalized by $\langle T^{\sigma l}_{\langle ij\rangle} \rangle $.  }
    \label{fig:square_times}
\end{figure}

In Fig.~\ref{fig:square_times}, we present additional $\tau^l$ values for square Hubbard models with PBC. The values of $\tau^l$ are also renormalized by the optimized $\langle T^{\sigma l}_{\langle ij\rangle}\rangle$ which are generally $O(-t)$. For closed shell systems at $U=4$ ($6\times 6$ with $n=2/3$, $8\times 8$ with $n=0.6875$), we see similar results to those of cylindrical pinned systems at low $U$, suggesting that similar results could be obtained with less slices $N_\ell$. 

On average these $|\tau^l|$ values are around $\approx 0.005-0.1$ and the corresponding Hubbard-Stratonvich coupling projection time (via $\lambda_i$) is of the same order, but can get as large as $\tau^l \approx 0.13$. 

\end{document}